\documentclass[preprint]{aastex}
\begin{document}

\title{The New Eclipsing Cataclysmic Variable SDSS 154453+2553
\footnote{Based on Observations obtained at the MDM Observatory operated by Dartmouth College, Columbia University, The Ohio State University, and the University of Michigan}
}

\author{Julie N. Skinner and John R. Thorstensen}
\affil{Dept. of Physics and Astronomy, 6127 Wilder Laboratory, Dartmouth College, Hanover, NH 03755-3528;\\
jns@dartmouth.edu}

\author{Eve Armstrong}
\affil{Dept. of Astronomy, Columbia University, 550 West 120th Street, New York , NY, 10027}

\and

\author{Steve Brady}
\affil{AAVSO, 25 Birch Street, Cambridge, MA, 02138}

\begin{abstract}
The cataclysmic variable SDSS154453+2553 was recently identified in the Sloan Digital Sky Survey.  We obtained spectra and photometry at the MDM Observatory, which revealed an eclipse with a 6.03 hour period.  The H$\alpha$ emission line exhibits a strong rotational disturbance during eclipse, indicating that it arises in an accretion disk.  A contribution from an M-type companion is also observed.  Time-series photometry during eclipse gives an ephemeris of 2454878.0062(15) + 0.251282(2)E. 
We present spectroscopy through the orbit and eclipse photometry.  Our analysis of the secondary star indicates a distance of $800 \pm 180$ pc.
\end{abstract}

\section{Introduction}
Cataclysmic variable stars (CVs) are close binary systems containing a white dwarf accreting matter from an extended secondary star through Roche-lobe overflow.  In systems with a low white dwarf magnetic field, an accretion disk can form.  The accretion disk is the origin for the strong emission lines characteristic of CV spectra not taken in outburst.  In many cases, the strong disk emission outshines the absorption features from the secondary star.   \citet{Warner} gives a comprehensive review of CVs.

Beyond observational difficulties, CVs (like most astronomical objects) present challenges in data interpretation.  The determination of binary system parameters suffers from many potential systematic effects.  Typically, radial velocities from emission lines originating in the accretion disk are used to constrain the mass ratio of the stars, but these features do not always directly map the motion of the white dwarf \citep{Wade85, MH90} in part because of asymmetry in the brightness of the disk and line broadening due to Keplerian motion of the disk.  When features from the secondary are observable, velocities may be calculated from its absorption lines.  While the absorption lines generally have a larger velocity amplitude than the emission lines, interpretation of absorption velocities is difficult as the secondary star is by no means a ``textbook" star \citep{Beu98, Knigge06}.  The donor star fills its Roche lobe and is therefore not spherical.  Its surface is being irradiated on one side, as well.   Despite these difficulties, eclipsing CVs are valuable because they provide the additional constraint on the inclination of the system.

Our CV of interest, SDSS 154453+2553, was discovered in the Sloan Digital Sky Survey \citep{SDSSDR7, Szkody09}.  Spectra from SDSS show single peaked emission lines with a detectable contribution from an M-dwarf \citep{Szkody09}, consistent with our data.  Photometry from SDSS gives a $g$ magnitude of 16.60.

\section{Observations}
Our data consist of  time-series photometry (see Section 2.1) and spectra (see Section 2.2).  Our early spectra showed signs of a rotational disturbance in the emission lines and follow-up time-series photometry soon confirmed that the system eclipses.  Radial velocity and eclipse observations taken in 2008 and 2009 were, when combined, sufficient to establish an unambiguous period of just over 6 hours (see below).
\subsection{Photometry}
\begin{figure}
\figurenum{1}
\includegraphics[angle = -90,width={\columnwidth}]{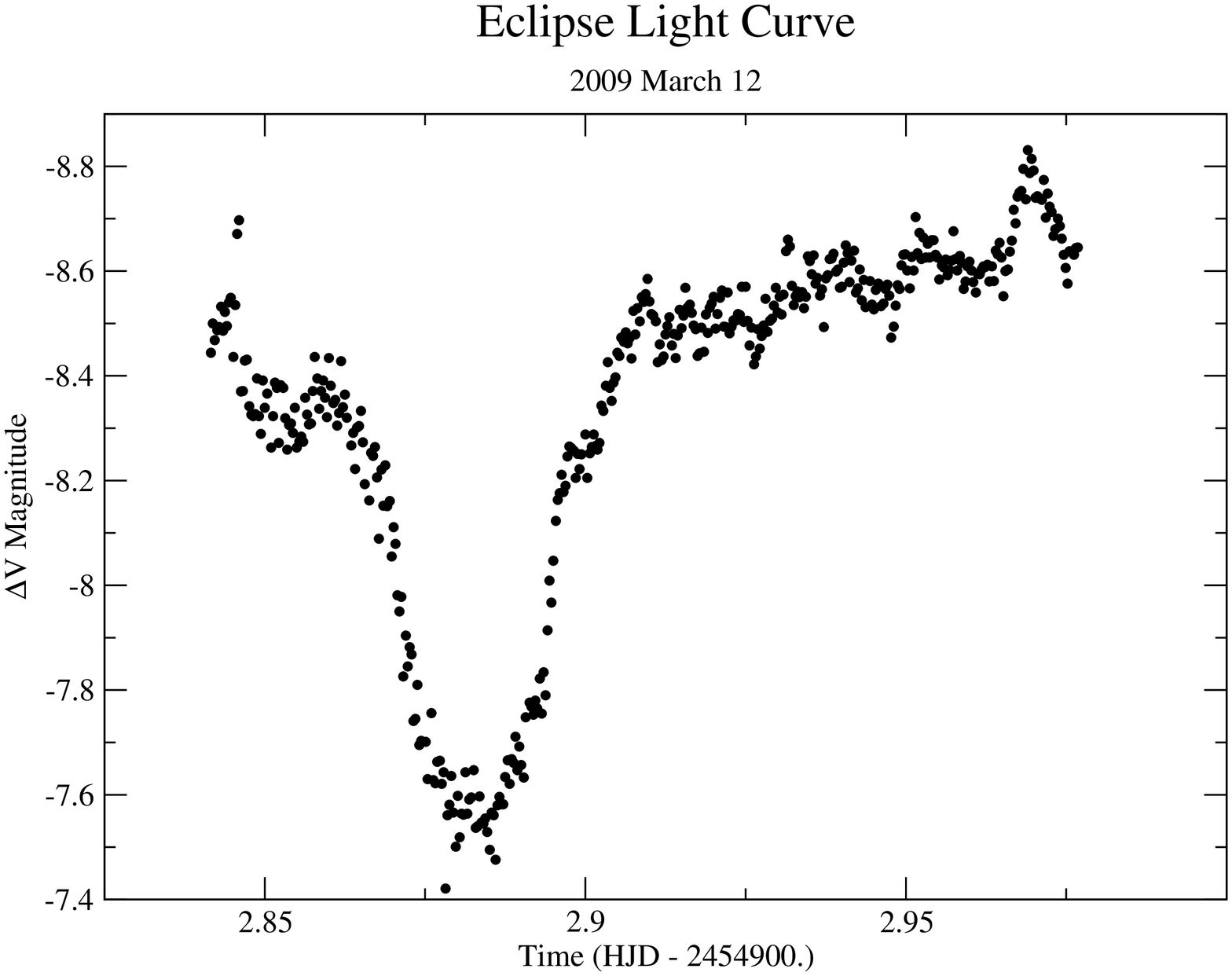}
\caption{The differential V light curve for SDSS 154453+2553.}
\label{lightcurve}
\end{figure}

Time-series photometry was taken in June 2008, February 2009 and March 2009 at the MDM Observatory 1.3m telescope using a SITe $1024^2$ CCD and binned 2x2.  Figure \ref{lightcurve} shows differential V photometry taken during a June 2008 run at the MDM 1.3m telescope.  Additional photometry, taken by Steve Brady with a 0.25 m robotic telescope in southern New Hampshire, was used along with the 1.3m data to determine an accurate ephemeris.  

The system showed small variations in brightness before and after eclipse with an eclipse depth of approximately one magnitude.  There is no clear indication of the time of bright spot eclipse in our light curve.  However, the feature at HJD 2454902.9 appears to be the white dwarf egress.  The photometry and spectroscopy establish the ephemeris to be 
\begin{equation}
\hbox{HJD mid-eclipse} = 2,454,878.0062(15) + 0.251282(2)E
\end{equation}
where E is the integer cycle count.  The eclipse width, $\Delta\phi = 0.11$ ($\sim$41 min.), was measured halfway down the steep sides of the light curve profile (Figure \ref{lightcurve}), which corresponds to the brightest part of the disk being covered and uncovered.

\subsection{Spectroscopy}
Spectra covering an entire orbit were taken in June 2008 with additional data taken in August 2008 and January 2009.  These were obtained using the 2.4m Hiltner telescope at the MDM Observatory using the modular spectrograph and a SITe $2048^2$ CCD, which gave a resolution of 3.5 \AA.  Our coverage was from 4210-7520 \AA~ with vignetting toward the ends.  Exposures were typically 300s in good seeing.  We performed our wavelength calibration using comparison lamp spectra taken during twilight and shifts from the 5577 \AA~ night sky line.  

\begin{figure}
\figurenum{2}
\includegraphics[angle = -90,width={\columnwidth}]{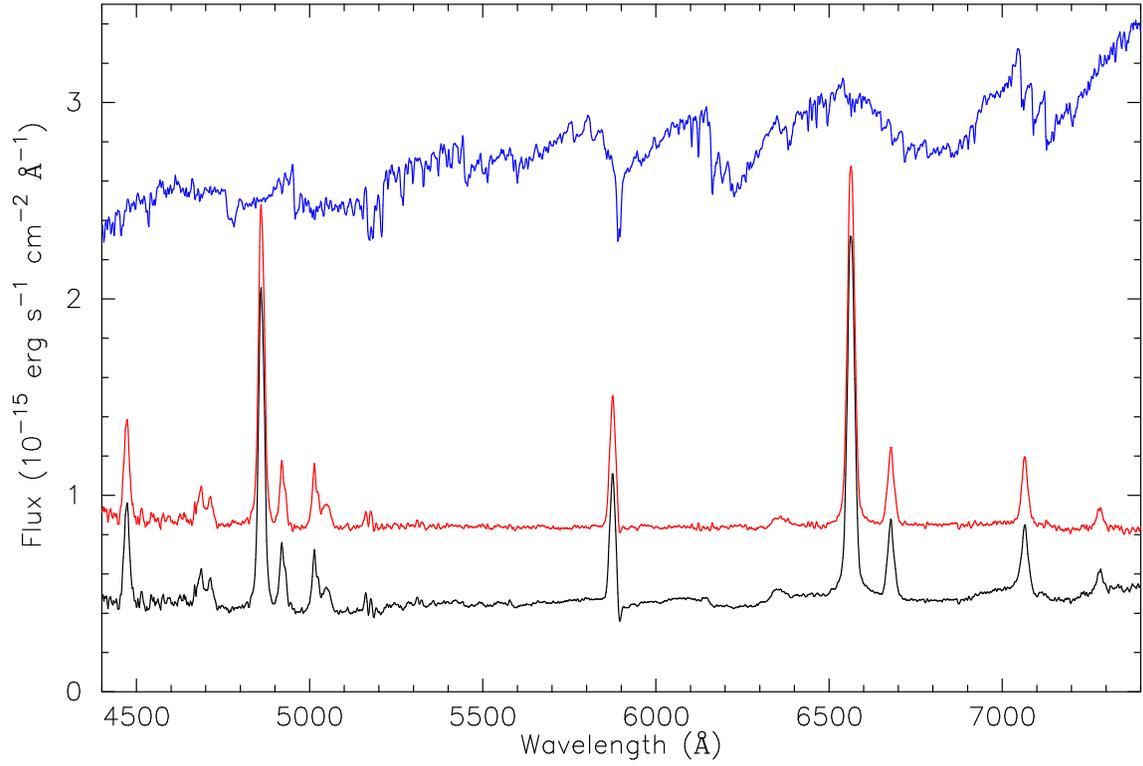}
\caption{A flux averaged spectrum in the rest frame of the secondary is plotted as the bottom spectrum.  The average spectrum shows single peaked lines with a weak contribution from an M-dwarf secondary.  The middle spectrum is the residual after subtracting a scaled M1.5 spectrum of Gliese 526 from our average spectrum, shown as the top spectrum.  }
\label{avgspec}
\end{figure}

The average spectrum shows emission lines as well as weak but detectable absorption from the secondary M-dwarf (top line, Figure 2).  The emission features appear single-peaked throughout the orbit.  To refine the spectral type of the secondary further, we subtracted a grid of M-dwarf spectra  \citep{Boeshaar76}  from our average spectrum and inspected the results by eye.  At the best fit spectral type, we should be left with only continuum and emission features, while the M-dwarf contribution should ideally disappear.  The best cancellation of the M-dwarf features occurs at type M$1.5 \pm 1.0$ (see Figure \ref{avgspec} ).    By comparing the contribution from the M-dwarf in SDSS1544 to the template spectra used in the decomposition (with known apparent magnitude), we can determine an apparent magnitude for the donor star of $m_V = 19.0 \pm 0.3$.  \citet{Knigge06} predicts a spectral type of K6.6 -- earlier than our determination.  Our later spectral type determination implies that the secondary is evolved.  \citet{Hoard01} found a similar result in the system CM Phe.  Although there was some ambiguity in their period determination, they determined a spectral type of M2-5 from a spectral deconvolution, while the predicted spectral type \citep{SmithDhillon98} is K4-M1.  The subject of late spectral types in CV donor stars is also discussed in \citet{BK00}.  For an orbital period close to 6 hours, their models predict possible spectral types of K4-M2 depending  primarily on q.  In addition, two stars (XY Ari \& LL Lyr) presented in \citet{Knigge06}'s Table 2 have orbital periods close to 6 hours with spectral type determinations of M$0 \pm 0.5$ and M$2.5 \pm 1.5$, respectively.  
  
Emission line velocities from H$\alpha$ were determined using a convolution method \citep{SY1980, Shafter83}.  In short, this method uses an antisymmetric function consisting of positive and negative Gaussians offset by an adjustable separation, $\alpha$, which is convolved with the observed line profile.  The zero of the convolution is taken as the line center.  We adopted a value of $\alpha = 2080$ km s$^{-1}$ for SDSS1544.  To determine this value, we used the method developed by \citet{Shafter83} to create a diagnostic diagram.  This procedure allowed us to determine the maximum separation of the Gaussians for which the signal-to-noise ratio was sufficient to probe the line wings. 

Velocities from the absorption component of the spectrum were determined using the \textit{xcsao} task in the \textit{rvsao} package of IRAF \citep{KM98}.  This was done by cross-correlating the spectrum between 5900 \AA~ and 6500 \AA~ with a composite M-dwarf spectrum with known radial velocities \citep{M87}. 

\begin{figure}
\figurenum{3}
\includegraphics[angle = -90,width={\columnwidth}]{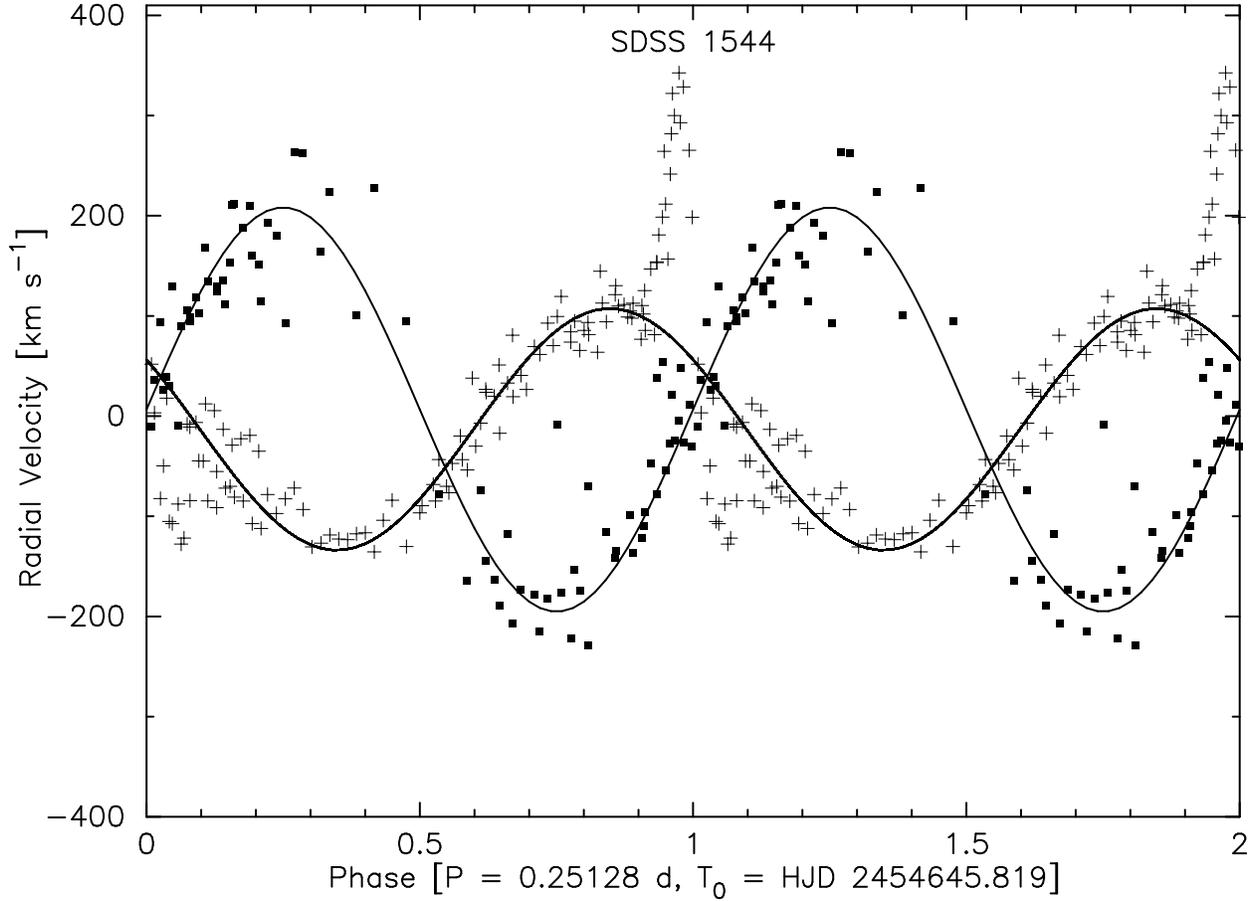}
\caption{Velocities from H$\alpha$ are shown by plus signs while M-dwarf absorption is shown as filled squares.  These are shown with their best-fit sinusoids as the solid lines.  The data are repeated for a second cycle for continuity.  The velocities measured during the eclipse are not included in the sine fit. The velocities derived from the H$\alpha$ emission are not exactly out of phase with the absorption velocities indicating asymmetric emission from the accretion disk.  There is also a clear rotational disturbance seen in the H$\alpha$ velocities.}
\label{rvcurve}
\end{figure}

Figure \ref{rvcurve} shows the measured velocities from emission and absorption features along with their respective best-fit sine curves. These sine curves have the form $v(t) = \gamma + K \sin[2\pi (t - T_o)/P]$ and Table 1 gives the fit parameters for each curve.  Emission velocities affected by the rotational disturbance were excluded from the sinusoidal fit.  If the emission lines were to represent the white dwarf motion, the eclipse should occur at $\phi = 0.5$ in the convention used here.  However, there is a phase offset of $\Delta\phi=0.087 \pm 0.007$.  This implies that the disk emission is somewhat asymmetric and these velocities are not faithfully representative of the motion of the white dwarf.  The difference in the mean velocity between the emission and absorption curves also indicates that one is affected by something other than the stellar motion. Another interesting feature seen in Figure \ref{rvcurve} is a clear rotational disturbance in the emission velocities seen as large deviations from the expected velocity.  This is most obvious at eclipse ingress.  A rotational disturbance occurs when the accretion disk is only partially eclipsed.  As the secondary star moves in front of the rotating disk, we see only the redshifted side of the disk instead of the integrated light.  During eclipse egress, the opposite phenomenon occurs and we observe only the blueshifted side.
 
\section{Analysis}
Normally, our observations would allow us to calculate the masses of the stars.  However, this is complicated by the fact that the emission line velocities do not represent the true motion of the white dwarf.   In addition, the absorption velocities show significant scatter with individual velocities having an uncertainty of 40 km s$^{-1}$.  This results in an error of 8 km s$^{-1}$ in the absorption velocity amplitude and a greater than 10\%  in the mass calculation even if systematics were negligible. Even though the emission line velocities do not accurately represent the motion of the white dwarf, we will calculate the mass ratio for illustrative reasons.  If we take the observed values as true indicators of the motion of the stars, we obtain $q  = K_1/K_2 = M_2/M_1 = 0.52$.  For a highly inclined object, this would imply a white dwarf mass of 0.45 $M_\sun$, which would give a companion mass of 0.23 $M_\sun$.  This value is slightly lower than the typical value for  an early M-dwarf star on the main sequence.  We should be clear, though, that these values are likely not correct since the emission velocities do not track the white dwarf motion.  However, this yields reasonable values for the masses of the stars and is a useful point of reference.  Combining $q=0.52$ and the eclipse half-width, we can constrain the inclination of the system to be $i=81^{\circ} \pm 1^{\circ}$.

We derive a distance to SDSS1544 through two different methods.  The first combines our apparent magnitude from the spectral decomposition with an absolute magnitude found using the relationship between surface brightness and spectral type for normal stars \citep{Beu99} (see Table \ref{Table 2}).  From \citet{BK00} evolutionary models, we can find a $R_2$ corresponding to our calculated mass for models with $P_{\rm{orb}} = 6.0$ hr.  Using their tabulated mass range for $M_2$ of 0.248-0.53 $M_\sun$, we find a radius range of 0.488-0.635 $R_\sun$.  Adopting $R_2 = 0.55 \pm 0.07 R_\sun$, we can convert surface brightness to $M_V$ and therefore find the corresponding $M_V$ for our M1.5 dwarf.  We corrected for extinction using the infrared dust maps of \citet{Sch98}.  These maps provide our best estimate for the amount of extinction due to dust.  However, the method is imperfect as they give the amount of extinction to the edge of the Galaxy rather than to SDSS1544.  Patchiness at scales below their 6-arcmin resolution would also affect the extinction estimate.  Combining our values for $m_V$, $M_V$, and $A_V$, we calculate a distance of $800 \pm 180$ pc.  We may also apply the donor sequence derived by \citet{Knigge06} to calculate a distance (see, again, Table \ref{Table 2}).  Using the K magnitude from the 2MASS survey and the absolute magnitude from \citet{Knigge06}'s empirical donor sequence, we get a distance of 810 pc.

\section{Conclusion}
We find that SDSS1544 is an eclipsing CV with a period of 6.03 hr, and determine an ephemeris.  The H$\alpha$ emission velocities show a pronounced rotational disturbance.  We also note that the determined emission velocities are not exactly out of phase with the absorption, which would point to asymmetric disk emission.  The decomposed spectrum is fitted best with a M1.5 star, which indicates that the secondary is somewhat evolved.  We also calculate the distance to SDSS1544 by two different methods.  Both result in a distance of roughly $800$ pc.

\section{Acknowledgements}
This work was supported by NSF grants AST-0307413 and AST-0708810.  The authors wish to thank the MDM staff for observing support, as well as Joe Patterson for donation of observing time.  We would also like to thank an anonymous referee for helpful comments and careful reading of the manuscript.

\begin{deluxetable}{ccccccc}
\tabletypesize{\small}
\setlength{\tabcolsep}{0.07in}
\tablecolumns{7}
\tablewidth{0in}
\tablecaption{Sinusoidal Fits}
\tablehead{
\colhead{Data Set} &
\colhead{Epoch (days)} &
\colhead{Period (days)} &
\colhead{K\tablenotemark{a} (km s$^{-1}$)} &
\colhead{$\gamma$\tablenotemark{a} (km s$^{-1}$)} &
\colhead{$\sigma$\tablenotemark{b} (km s$^{-1}$)} &
\colhead{N}
}
\startdata
H$\alpha$ emission & 54647.724(3) & 0.251282 & 101(7) & $-44(5)$ & 34 & 121\\
Absorption & 54645.8182(15) &         0.251282 &  193(8) & $ 11(6)$ & 40 &  78 \\ 
\enddata
\tablenotetext{a}{K is the amplitude of the sinusoidal fit and $\gamma$ is the mean velocity.}
\tablenotetext{b}{Error in a single measure based on the scatter around the best fit}
\end{deluxetable}

\begin{deluxetable}{ccccc}
\tabletypesize{\small}
\setlength{\tabcolsep}{0.07in}
\tablecolumns{5}
\tablewidth{0in}
\tablecaption{Values used in Distance Calculations}
\tablehead{
\colhead{Method} &
\colhead{Absolute Magnitude} &
\colhead{Apparent Magnitude} &
\colhead{Extinction} &
\colhead{Distance (pc)} 
}
\startdata
\citet{Beu99} & 9.29 $\pm$ 0.58(V) & 19.0 $\pm$ 0.3(V) & 0.2 $\pm$ 0.1 & 800 $\pm$ 180\\
\citet{Knigge06} & 4.48(K) & 14.019 $\pm$ 0.065(K) & --- & 810\\ 
\enddata
\label{Table 2}
\end{deluxetable}

\end{document}